\documentclass[aps,twocolumn,superscriptaddress,preprintnumbers,pre]{revtex4-1}
\usepackage{xr-hyper}
\usepackage{mathrsfs, hyperref}
\usepackage{amssymb, amsbsy, amsmath, latexsym, dsfont, array, layout, graphics,mathrsfs,braket,amsfonts,amsthm,
  amssymb,graphicx,subfigure, youngtab,color,bm,mathtools,braket,verbatim,url,cleveref,natbib,hypernat,extarrows,inputenc,multirow,bbm,dsfont}
  
\usepackage{xcolor}
\colorlet{RED}{red}
\usepackage{graphicx,psfrag}
\usepackage{amsfonts}
\usepackage[figuresright]{rotating}  
\usepackage{amssymb}
\usepackage{amsmath}
\usepackage{subfigure}
\usepackage{multirow}
\usepackage{tabularx}
\usepackage[resetlabels]{multibib}
\usepackage{amssymb, amsbsy, amsmath, latexsym, dsfont, array, layout, graphics,mathrsfs,braket,amsfonts,amsthm,
  amssymb,graphicx,subfigure,dcolumn, youngtab,color,bm,mathtools,braket,verbatim,url,cleveref,natbib,hypernat,extarrows,inputenc,multirow}
\usepackage{xcolor}
\graphicspath{ {./images/} }

\usepackage{mathrsfs, hyperref}
\usepackage{amssymb, amsbsy, amsmath, latexsym, dsfont, array, layout, graphics,mathrsfs,braket,amsfonts,amsthm,
  amssymb,graphicx,subfigure, youngtab,color,bm,mathtools,braket,verbatim,url,cleveref,natbib,hypernat,extarrows,inputenc,multirow,bbm}
\usepackage{graphicx}
\usepackage{dcolumn}
\usepackage{bm}
\usepackage{xcolor}
\usepackage[normalem]{ulem}
\usepackage[T1]{fontenc}
\graphicspath{ {./images/} }
\allowdisplaybreaks 

\newcommand{\splitatcommas}[1]{%
\begingroup 
\begingroup\lccode`~=`, \lowercase{\endgroup 
\edef~{\mathchar\the\mathcode`, \penalty0 \noexpand\hspace{0pt plus 1em}}%
}\mathcode`,="8000 #1%
\endgroup 
} 

\begin{document}
\title{Ideal topological flat bands in chiral symmetric moir\'e systems from non-holomorphic functions}
\author{Siddhartha Sarkar}\thanks{These three authors contributed equally}
\affiliation{%
Department of Physics, University of Michigan, Ann Arbor, MI 48109, USA
}
\author{Xiaohan Wan}\thanks{These three authors contributed equally}
\affiliation{%
Department of Physics, University of Michigan, Ann Arbor, MI 48109, USA
}
\author{Yitong Zhang}\thanks{These three authors contributed equally}
\affiliation{%
Department of Physics, University of Michigan, Ann Arbor, MI 48109, USA
}
\author{Kai Sun}
\email{sunkai@umich.edu}
\affiliation{%
Department of Physics, University of Michigan, Ann Arbor, MI 48109, USA
}
\begin{abstract}
Recent studies on topological flat bands and their fractional states have revealed increasing similarities between moir\'e flat bands and Landau levels (LLs). For instance, like the lowest LL, topological exact flat bands with ideal quantum geometry can be constructed using the same holomorphic function structure, $\psi_{\mathbf{k}} = f_{\mathbf{k}-\mathbf{k}_0}(z) \psi_{\mathbf{k}_0}$, where $f_{\mathbf{k}}(z)$ is a holomorphic function. This holomorphic structure has been the foundation of existing knowledge on constructing ideal topological flat bands. In this Letter, we report a new family of ideal topological flat bands where the $f$ function does not need to be holomorphic. We provide both model examples and universal principles, as well as an analytic method to construct the wavefunctions of these flat bands, revealing their universal properties, including ideal quantum geometry and a Chern number of $C = \pm 2$ or higher.
\end{abstract}
\maketitle

Topological flat bands have long been a central focus in modern condensed matter physics due to their unique physical properties and the novel quantum phenomena they can host, such as the fractional quantum Hall effect~\cite{landau2013quantum, hofstadter1976energy, tang2011high,sun2011nearly, neupert2011fractional, sheng2011fractional, regnault2011fractional}. Recent research has revealed two intriguing insights about these bands: (1) Besides Landau levels (LLs), topological flat bands and the fractional states that they may host can be realized in various physical systems, such as moir\'e systems, even in the absence of external magnetic fields~\cite{cai2023signatures, zeng2023thermodynamic, park2023observation, xu2023observation,lu2024fractional}. (2) Although these moir\'e topological flat bands are created in fundamentally different setups from LLs, they appear to share the same theoretical foundation.

To illustrate the deep connection between moir\'e flat bands and LLs, we consider the lowest LL and chiral twisted bilayer graphene (TBG) as examples. Unlike the lowest LL, which exhibit perfect band flatness and ideal quantum geometry, flat bands in moir\'e systems are typically not ideal—they are not perfectly flat and their wavefunctions lack ideal quantum geometry. However, these systems can often be adjusted slightly to make their topological flat bands ideal. A well-known example is the chiral limit of TBG~\cite{tarnopolsky2019origin, wang2021chiral}. Similar ideal topological flat bands can also arise in a variety of systems, such as twisted bilayer checkerboard lattices~\cite{li2022magic}, TBG with a spatially alternating magnetic field~\cite{le2024double}, single-layer systems with quadratic band crossing points under a periodic strain field~\cite{wan2023topological, sarkar2023symmetry}, the ideal limit of twisted bilayer Fe-based superlattices~\cite{eugenio2023twisted}, TBG with strong second harmonic tunneling~\cite{becker2023degenerate}, and chiral twisted bilayer systems with higher-order topological nodes~\cite{cui2024classification}.

For LLs, the foundation of our understanding are based on a simple fact: due to the algebra of magnetic translations, the lowest Landau level exhibits a simple structure, $\psi = f(z) \exp(-\frac{z\bar{z}}{4l^2})$, where $l$ is the magnetic length and $z=x+i y$ is the complex coordinate. Up to a less important factor, the eigenstates are holomorphic functions of $z$. This holomorphic structure is crucial for understanding both the integer and fractional quantum Hall effects, enabling us to easily formulate their wavefunctions, such as the Laughlin wavefunction, and understand their physical properties. In chiral TBG, as well as other ideal flat band models mentioned above, wavefunctions of the exact flat bands exhibit a structure identical to the lowest
LL $\psi_\mathbf{k}(\mathbf{r}) = f_{\mathbf{k} - \mathbf{K}}(z) \psi_\mathbf{K}(\mathbf{r})$~\cite{tarnopolsky2019origin}, where $f_{\mathbf{k} - \mathbf{K}}(z)$ is a holomorphic function and $\mathbf{K}$ represents the corner of the moir\'e Brillouin zone. 
More importantly, the analogy to the lowest LL extends even into the strongly correlated regime. Similar to Landau level systems~\cite{haldane1985periodic}, we can derive exact solutions for fractional states from such ideal exact flat bands~\cite{ledwith2020fractional, wang2021exact, ledwith2023vortexability}.

It is worthwhile to mention that the deep connection between moir\'e exact flat bands and LLs extends beyond just the simplest cases. In Landau level systems, more complex exact flat bands and fractional states can be achieved by introducing additional elements to the lowest Landau level. For example, higher LLs can be obtained by applying the raising operator $2\partial_z -\frac{\overline{z}}{2l}$, and quantum Hall multilayers can be created by introducing additional layers. In moir\'e systems, by adding more components, layers, or bands, higher vortexability~\cite{fujimoto2024higher,liu2024theorygeneralizedlandaulevels} and exact flat bands with higher Chern numbers~\cite{ledwith2022family,wang2022hierarchy} can be achieved.

Despite recent studies increasingly revealing evidence of deep and fundamental connections between moir\'e flat bands and LLs,
this Letter explores potential differences between LLs and moir\'e exact flat bands. We report a new family of exactly flat topological bands that are vortexable~\cite{ledwith2023vortexability} but cannot be directly traced back to the lowest Landau level. Specifically, we find that in these flat band systems, the Bloch wavefunction can be written as $\psi_{\mathbf{k}} = f_\mathbf{k-\mathbf{k}_0}(z, \bar{z}) \psi_{\mathbf{k}_0}$, where $\psi_{\mathbf{k}_0}$ is the wavefunction of the same flat band at an arbitrarily chosen reference momentum point $\mathbf{k}_0$. In contrast to the lowest LL or chiral TBG, where $f$ must be a holomorphic function, here this $f$ function is not holomorphic.

This distinction has two important consequences for moir\'e flat bands. First, because the wavefunction needs to obey the Bloch boundary conditions, if the function $f$ is holomorphic, it must have a pole in the moir\'e unit cell. As a result, to support ideal flat bands, the wavefunction $\psi_{\mathbf{k}_0}$ must have at least one zero to cancel the divergence at this pole. In our system, because $f$ is not holomorphic, it does not need to have any poles, and thus $\psi_{\mathbf{k}_0}$ does not need zeros. Secondly, for holomorphic functions of $f$, the holomorphic nature implies that the Chern number must be $C = \pm 1$, unless additional degrees of freedom (e.g., using multiple layers~\cite{ledwith2022family,wang2022hierarchy} or models that can be re-written as two decoupled $C=1$ flat bands~\cite{li2022magic,sarkar2023symmetry,herzog2024topological}). However, the flat bands we find here are not confined by this limit. In contrast, they generally have a Chern number of $C= \pm 2$ or higher.

Beyond specific model examples, we further demonstrate that these instances are not isolated cases but represent one of two possible pathways toward ideal moir\'e flat bands: 
besides the well-known holomorphic option, $f(z)$, there exists a second viable pathway where $f$ is non-holomorphic. This second option is prohibited in LLs or systems that can be mapped to Landau-level-like structures, but it is possible for general moir\'e systems.
We further derive the partial differential equations (PDEs) that a non-holomorphic $f$ function must satisfy. Interestingly, these equations can be solved analytically, with solutions expressible as integrals of theta functions. From these analytic solutions, it can be proven that these bands must have ideal quantum geometry and Chern numbers $C=\pm 2$.

In the discussion section, we delve into the impact of this new knowledge, revealing that the $f$ function solution that we identify has an interesting connection to the ratio between wavefunctions from the first and lowest LLs. We will also comment on the potential implications for fractional quantum states.

\begin{figure}[t]
     \centering
\includegraphics[width=0.4 \textwidth]{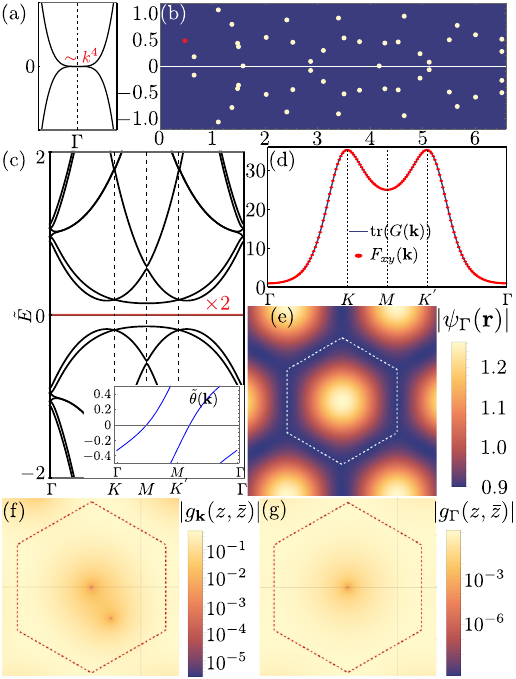}
     \caption{
     Exact flat bands from a quartic band crossing point under a moir\'e potential with $p31m$ symmetry. (a) Band structure of a quartic band crossing point. (b) Magic values of $\alpha$. Each point represent a magic $\alpha$. For better visualization, we take a power $1/4$ for the values of $\alpha$. The horizontal and vertical axis are the real and imaginary part of $\alpha^{1/4}$, respectively.  
     (c) Band structure along the high symmetry path in the Moir\'e Brillouin zone (MBZ). The potential $A(\mathbf{r})$ is shown in Eq.~\ref{eq:p31m} and $\alpha=-0.209$ [the red dot in (b)]. The vertical axis is normalized eigen-energies $\tilde{E}=\frac{E}{|\mathbf{G}_{i}|^2}$. There are 2 exact flat bands at energy $\tilde{E}=0$. Wilson loop spectrum $\tilde{\theta}({\mathbf{k}})=\frac{\theta({\mathbf{k}})}{2\pi}$ of the sublattice polarized wavefunction is shown at the right bottom corner of (c), which implies Chern number $|C| = 2$.
     (d) The trace of Fubini-Study metric $G(\mathbf{k})$ and the Berry curvature $F_{xy}(\mathbf{k})$ for one sublattice polarized band along the high symmetry path in the MBZ. These two quantity perfect agree with each other, i.e., ideal quantum geometry. 
     (e) Density plot of $|\psi_\Gamma(\mathbf{r})|$ which has no zeros in the unit cell.
     (f) Density plot of $|g_\mathbf{k}(z,\bar{z})|$ at $\mathbf{k}=(0.21,0.13)$, which has zeros at both the origin and $\frac{\sqrt{3} a^2}{4 \pi}(0.13,-0.21)$.
     (g) Density plot of $|g_\Gamma(z,\bar{z})|$ which has one zero at the origin. Dashed lines in (e,f,g) mark the boundary of a unit cell.
     }
     \label{fig:ExactFlatBand}
\end{figure}

\textit{A simple example}--
We begin our discussion by examining a simple example that illustrates the key properties of this new family of exact flat bands. For demonstration purposes, we select a model whose Hamiltonian and eigenwavefunctions take the simplest forms, rather than focusing on identifying the most feasible setup for future experimental realization. However, it is important to emphasize that although this model is intended for demonstration purposes, the physics it showcases are generic and represent one of the two possible pathways towards ideal topological flat bands, as will be shown in subsequent sections.

Consider a $2\times 2$ Hamiltonian:
\begin{equation}
\label{eq:H4}
    \mathcal{H}_4(\mathbf{r}) = \begin{pmatrix}
        0 & \mathcal{D}_4^\dagger(\mathbf{r})\\
        \mathcal{D}_4(\mathbf{r}) & 0
    \end{pmatrix}, \mathcal{D}_4(\mathbf{r}) = (2i\overline{\partial_z})^4+\alpha A(\mathbf{r}),
\end{equation}
where $\partial_z = \frac{1}{2}(\partial_x - i\partial_y)$, the overbar indicates complex conjugation, $A(\mathbf{r})$ is a periodic moir\'e potential, and $\alpha$ is the complex amplitude of this potential. In the absence of the moir\'e structure ($\alpha = 0$), this Hamiltonian represents a 2D system with a quartic band crossing, where the dispersion near the band crossing follows $E \propto |\mathbf{k}|^4$ {\color{black}(Fig.~\ref{fig:ExactFlatBand}(a))}. The moir\'e potential $\alpha A(\mathbf{r})$ can be induced by a periodic strain ( quadrupole) or 16-pole field, which may result from lattice mismatch with the substrate or moir\'e lattice reconstruction. This Hamiltonian is both chiral symmetric $\{\mathcal{H}_4,\mathcal{S}\} = 0$ with $\mathcal{S} = \sigma_z$ and time-reversal symmetric $[\mathcal{H}_4,\mathcal{T}] = 0$ with $\mathcal{T} = \sigma_x K$ and $K$ being complex conjugation. For $A(\mathbf{r})$, we require it to preserve three-fold rotational symmetry $\mathcal{C}_{3z}$. Given that $(\overline{\partial_z})^4 \rightarrow e^{2\pi i/3}(\overline{\partial_z})^4$ under a three-fold rotation, the moir\'e potential must obey $A(\mathcal{C}_{3z}\mathbf{r}) = e^{2\pi i/3}A(\mathbf{r})$. As will be discussed below and in the Supplementary Material (SM)~\cite{SM2024}, the combination of these symmetries and the fact that $\mathcal{D}_4$ only has antiholomorphic derivative ensures that the quartic band crossing does not split into band crossings of lower order (e.g., Dirac or quadratic band crossings).

Here we choose a simple moir\'e potential with only first harmonics
\begin{equation}
    A(\mathbf{r}) = \frac{1}{2}\sum_{i=1}^3 e^{2\pi i (n-1)/3} e^{-i\mathbf{G}_i\cdot\mathbf{r}},
    \label{eq:p31m}
\end{equation}
where $\mathbf{G}_i$ represents the reciprocal lattice vector of the moir\'e structure $\mathbf{G}_i= \frac{4\pi}{\sqrt{3}a}(-\sin(\frac{2\pi}{3}(i-1)),\cos(\frac{2\pi}{3}(i-1))$ with $a$ being the moir\'e period. Note that for this $A(\mathbf{r})$, a mirror symmetry $\mathcal{M}_y$, $(x,y) \to (x,-y)$, emerges when $\alpha$ is real. To find if this model support exact flat bands at certain ``magic'' values of $\alpha$, we utilize a method introduced in Ref.~\cite{becker2022mathematics}. 
Here we construct the Birman-Schwinger operator~\cite{becker2022mathematics, becker2021spectral}
\begin{equation}
    T_4(\mathbf{k};\mathbf{r}) = - (2i\overline{\partial_z}-k)^{-4}A(\mathbf{r}).
\end{equation}
where $k=k_x+ik_y$ is an arbitrary wavevector, and compute the eigenvalues of this operator $\eta_\mathbf{k}$. In {\color{black}Fig.~\ref{fig:ExactFlatBand}(b)} we plot the inverse of these eigenvalues $1/\eta_\mathbf{k}$ at a non-special $\mathbf{k}$, and numerically verified that these values are independent of $\mathbf{k}$. 
Each of these eigenvalues provides a ``magic" value of $\alpha=1/\eta_\mathbf{k}$,  at which exact flat bands emerge (see SM for details). In {\color{black}Fig.~\ref{fig:ExactFlatBand}(c)}, we plot the band structure of $\mathcal{H}_4(\mathbf{r})$ for one of these magic $\alpha$ values. As shown, there are two exact flat bands at zero energy. The wavefunctions of these two flat bands are sublattice-polarized, $\Psi_{\mathbf{k},1}(\mathbf{r})=\{\psi_\mathbf{k}(\mathbf{r}),0\}$ and $\Psi_{\mathbf{k},2}(\mathbf{r})= \mathcal{T}\Psi_{-\mathbf{k},1}(\mathbf{r})=\{0,\psi_{-\mathbf{k}}^*(\mathbf{r})\}$, where $\psi_\mathbf{k}(\mathbf{r})$ is a zero mode of $\mathcal{D}_{4}(\mathbf{r})$. It is straightforward to verify that 
$\Psi_{\mathbf{k},1}$ and $\Psi_{\mathbf{k},2}$ are related by time-reversal transformation and thus must carry opposite Chern numbers. Using the Wilson loop winding number [in the inset of Fig.~\ref{fig:ExactFlatBand}(c)], we determine their Chern numbers to be $\pm 2$. More interestingly, these two flat bands exhibit ideal quantum geometry $\text{tr}(G(\mathbf{k})) =|F_{xy}(\mathbf{k})|$, where $G(\mathbf{k})$ and and $F_{xy}(\mathbf{k})$ are the Fubini-Study metric~\cite{roy2014band, ledwith2020fractional, ledwith2021strong, wang2021exact, mera2021engineering, ledwith2022family} and the the Abelian Berry curvature respectively. 

{\color{black}Fig.~\ref{fig:ExactFlatBand}(e)} shows the wavefunction $\psi_\Gamma(\mathbf{r})$ at the $\Gamma$ point ($\mathbf{k} =\mathbf{0}$). One key feature to highlight is that $\psi_\Gamma(\mathbf{r})$ never reaches zero.  This feature directly indicates that $f_\mathbf{k}(z,\bar{z})=\psi_\mathbf{k}/\psi_\Gamma$ must not be an holomorphic function, in direct contrast to other ideal flat bands such as the lowest Landau level or chiral TBG. As will be discussed below, these two cases, $f_\mathbf{k}$ being holomorphic or non-homophobic, represent the two allowed pathways toward achieving ideal flat bands in moir\'e systems.

\textit{Construction of wavefunction.}--Although $f_\mathbf{k}$ is not holomorphic, we can still analytically construct Bloch wavefunctions for these ideal flat bands. The clue comes from {\color{black}Fig.~\ref{fig:ExactFlatBand}(f)}, where we plot the function $g_\mathbf{k}(z,\bar{z}) \equiv \overline{\partial_z}f_\mathbf{k}(z,\bar{z}) = \overline{\partial_z}(\psi_\mathbf{k}(\mathbf{r})/\psi_\Gamma(\mathbf{r}))$. This function shares three key properties with the Bloch wavefunction of ideal flat bands in chiral TBG: (1) Bloch periodic, $g(\mathbf{r}+\mathbf{a})=g(\mathbf{r}) e^{i \mathbf{k} \cdot \mathbf{a}}$, (2) having isolated zeros in the unit cell {\color{black}[Fig.~\ref{fig:ExactFlatBand}(f)]}, and most importantly, (3) ``vortexable."

To prove its vortexability, we begin with the equations that $\psi$ must satisfy, $\mathcal{D}_4\psi_\mathbf{k}(\mathbf{r}) = \mathcal{D}_4(f_\mathbf{k}(z,\bar{z}) \psi_\Gamma(\mathbf{r})) = 0$ and $\mathcal{D}_4\psi_\Gamma(\mathbf{r}) = 0$. By subtracting these two equations, we obtain the equation for $g_\mathbf{k}$
\begin{equation}
\label{Eq:geq}
\begin{split}
    \tilde{D}_4 g_\mathbf{k} \equiv \left(\sum_{n=0}^3 \binom{4}{n} (\overline{\partial_z}^n \psi_\Gamma)\overline{\partial_z}^{3-n}\right)g_\mathbf{k} = 0,
\end{split}
\end{equation}
where $\binom{n}{k} = \frac{n!}{k! (n-k)!}$ is the binomial coefficient. 
Because this differential equation does not contain $\partial_z$, if $g_\Gamma$ is a solution, then for any arbitrary holomorphic function $h(z)$, $h(z) g_\Gamma$ must also be a solution.

With these three properties of $g$, we can construct the $g$ function in the same manner as how wavefunctions in chiral TBG are constructed. First, 
by solving Eq.\eqref{Eq:geq} at  $\Gamma$ ($\mathbf{k} =\mathbf{0}$), we find a non-trivial solution for $g_\Gamma$.  Unlike the lowest Landau levels or the chiral limit of TBG, where $g$ is strictly zero, our $g_\Gamma$ only have one isolated zero at the center of the unit cell [{\color{black}Fig.~\ref{fig:ExactFlatBand}(g)}]. Around this zero, symmetry requires $g$ to obey the asymptotic form $g_\Gamma(z,\bar{z}) \propto z\bar{z}$,  which we have also verified numerically. This asymptotic form is crucial as it indicates that we can use this zero to cancel singularities caused by a pole. Hence, we can write down the function $g_\mathbf{k}$ as
\begin{equation}
\label{eq:gk}
\begin{split}
    g_\mathbf{k}(z,\bar{z}) &= h_\mathbf{k}(z;z_0) g_\Gamma(z,\bar{z}),\\
    h_\mathbf{k}(z;z_0) &=e^{\frac{i}{2} (\bar{k}z+k\bar{a}_1 z/a_1)}\frac{\vartheta\left(\frac{z-z_0}{a_1}-\frac{k}{b_2},\tau\right)}{\vartheta\left(\frac{z-z_0}{a_1},\tau\right)} ,
\end{split}
\end{equation}
where $a_1 = (\mathbf{a}_{1})_x+i(\mathbf{a}_{1})_y$ is the complexified moir\'e lattice vector $\mathbf{a}_1 = a(1,0)$, $b_2 = (\mathbf{b}_{2})_x+i(\mathbf{b}_{2})_y$ is the complexified moir\'e reciprocal lattice vector $\mathbf{b}_2 = 4\pi/\sqrt{3}a(0,1)$, $k = k_x+ik_y$, $\tau = e^{2\pi i/3}$, $\vartheta(z,\tau) = -i\sum_{n=-\infty}^\infty(-1)^n e^{\pi i \tau(n+1/2)^2+\pi i(2n+1)z}$ is the Jacobi theta function~\cite{TataI,ledwith2020fractional}, and $z_0 = 0$ is the position of the zero in $g_\Gamma(\mathbf{r})$. Since, $h_\mathbf{k}(z;z_0)$ is a holomorphic function of $z$, $g_\mathbf{k}$ satisfies Eq.~\eqref{Eq:geq}. From the definition of the theta function, it can be verified that the function $h_\mathbf{k}(z;z_0)$ is Bloch periodic. Furthermore, since $\vartheta(z,\tau)$ has simple zeros at positions $z=m_1+m_2\tau$, $m_i\in \mathds{Z}$, $h_\mathbf{k}(z;z_0)$ has a simple zero at $z=z_0+a_1 k/b_2$ and a simple pole at $z = 0$ in the unit cell. This and the fact that near $\mathbf{r} = \mathbf{0}$, $g_\Gamma(z,\bar{z}) \sim z\bar{z}$ implies that $g_\mathbf{k}(z,\bar{z})$ has a zero at $\mathbf{r} = \mathbf{0}$ (near which the function has the form $g_\mathbf{k}(z,\bar{z})\sim \bar{z}$), and another zero at $\mathbf{r}_\mathbf{k} = -\frac{\sqrt{3}a^2}{4\pi}\hat{z}\times\mathbf{k}$ (near which the function has the form $g_\mathbf{k}(\mathbf{r}_\mathbf{k}+(x,y))\sim z$). These two zeros can be seen in Fig.~\ref{fig:ExactFlatBand}(f). Constructing the function $g_\mathbf{k}(z,\bar{z})$ using Eq.~\eqref{eq:gk}, we numerically verified that it matches with the numerically solved function $g_\mathbf{k}(z,\bar{z})$ up to a constant factor. Note that the function $g_\mathbf{k}(z,\bar{z})$ carries Chern number $C = 1$ since it has the same form as the lowest LL wavefunctions on a torus~\cite{haldane1985periodic}. Also, note that the periodic part of $g_\mathbf{k}(z,\bar{z})$, which is $e^{-i\mathbf{k}\cdot\mathbf{r}}g_\mathbf{k}(z,\bar{z})$, is a holomorphic function of $k=k_x+ik_y$~\cite{ledwith2020fractional,sarkar2023symmetry}.

To obtain $f_\mathbf{k}(z,\bar{z})$, we can utlize the fact that $g_\mathbf{k}$ is a Bloch periodic function, and hence it can be written as a Fourier series $g_\mathbf{k}(z,\bar{z}) = \sum_\mathbf{G} g_\mathbf{k}(\mathbf{G}) e^{i(\mathbf{G}+\mathbf{k})\cdot\mathbf{r}}$, where $\mathbf{G} = m_1\mathbf{G}_1+m_2\mathbf{G}_2, m_i\in\mathds{Z}$, are the reciprocal lattice vectors and $g_\mathbf{k}(\mathbf{G})$ are the Fourier amplitudes. Because $\overline{\partial_z}f_\mathbf{k} = g_\mathbf{k}$, we get
\begin{equation}
\label{eq:fk1}
    f_\mathbf{k}(z,\bar{z}) = c(\mathbf{k}) \int d\bar{z}g_\mathbf{k}(z,\bar{z}) = c(\mathbf{k}) \sum_\mathbf{G} \frac{2g_\mathbf{k}(\mathbf{G})}{i(k+G)} e^{i(\mathbf{G}+\mathbf{k}).\mathbf{r}},
\end{equation}
where $k+G = (k_x+G_x)+i(k_y+G_y)$, and $c(\mathbf{k})$ is a $\mathbf{k}$ dependent constant which does not alter the form of the wavefunction at a given $\mathbf{k}$. The problem with this form $f_\mathbf{k}$ is that the summand at $\mathbf{G} = 0$ has a $1/k$ type singularity near $k=0$ (if $g_{\mathbf{k}=\mathbf{0}}(\mathbf{G}=\mathbf{0})\neq 0$, which we numerically verified). This fact and the fact that $f_{\mathbf{k} = 0} \equiv f_\Gamma = 1$, implies that to have a smooth gauge for $f_\mathbf{k}(z,\bar{z})$ near $\mathbf{k}=\mathbf{0}$, $c(\mathbf{k}) = ik/(2g_{\mathbf{k}=\mathbf{0}}(\mathbf{G}=\mathbf{0}))$. Hence, the full expression for $f_\mathbf{k}$ is
\begin{equation}
\label{eq:fk}
    f_\mathbf{k}(z,\bar{z})
    = \frac{k}{g_{\mathbf{k}=\mathbf{0}}(\mathbf{G}=\mathbf{0})} \sum_\mathbf{G} \frac{g_\mathbf{k}(\mathbf{G})}{k+G} e^{i(\mathbf{G}+\mathbf{k}).\mathbf{r}}
\end{equation}
This extra holomorphic factor $k=k_x+ik_y$ smooths out the gauge near $k=0$, but gives an extra winding to the Berry phase of $f_\mathbf{k}$ at the edge of the Brillouin zone. This extra winding, in addition to $g_\mathbf{k}$ having Chern number $C = 1$, implies that $f_\mathbf{k}$ carries Chern number $C = 2$; as was seen from the Wilson loop spectrum in the inset of Fig.~\ref{fig:ExactFlatBand}(c). Since $c(\mathbf{k})$ and the periodic part of $g_\mathbf{k}(z,\bar{z})$ are holomorphic in $k=k_x+ik_y$, the periodic part $u_\mathbf{k}(\mathbf{r}) = e^{-i\mathbf{k}\cdot\mathbf{r}}\psi_\mathbf{k}(\mathbf{r}) = e^{-i\mathbf{k}\cdot\mathbf{r}}f_\mathbf{k}(z,\bar{z})\psi_\Gamma(\mathbf{r})$ of the wavefunction $\psi_\mathbf{k}(\mathbf{r})$ is holomorphic in $k=k_x+ik_y$, which immediately implies ideal quantum geometry~\cite{claassen2015position,ledwith2020fractional}. 

It is worth noting that there is a similarity between the wavefunction we constructed above and the wavefunction for the exact flat bands in twisted mono-bilayer graphene. We can write the coupled set of equations satisfied by $f_\mathbf{k}$ and $g_\mathbf{k}$ in the following form
\begin{equation}
	\begin{pmatrix}
		\overline{\partial_z} & -1\\
		0 & \tilde{D}_4
	\end{pmatrix}\begin{pmatrix}f_\mathbf{k}\\g_\mathbf{k}\end{pmatrix} = \begin{pmatrix}0 \\0\end{pmatrix},
\end{equation}
where $\tilde{D}$ was defined in Eq.~\eqref{Eq:geq}. This equation has exactly the same form as the equation satisfied by the sublattice polarized wavefunction in twisted mono-bilayer graphene~\cite{ledwith2022family}:
\begin{equation}
	\begin{pmatrix}
		\overline{\partial_z} & -\beta\\
		0 & D_\text{TBG}
  \end{pmatrix}\begin{pmatrix}\psi_{\mathbf{k},1}\\\psi_{\mathbf{k},\text{TBG}}\end{pmatrix} = \begin{pmatrix}0 \\0\end{pmatrix}.
\end{equation}
This is why our flat band has Chern number $C=2$, same as the twisted mono-bilayer graphene. However, a crucial difference between these two systems is that, in our case, the lowest LL-type function $g_\mathbf{k}$ is an auxiliary function that does not correspond to an actual physical degree of freedom. In contrast, in the twisted mono-bilayer graphene system, the lowest LL-type function $\psi_{\mathbf{k}, \text{TBG}}$ represents a physical degree of freedom.

\textit{Generic origin of these ideal flat bands}--
In this section, we consider generic situations, focusing on the fundamental origins and essential ingredients of this new family of ideal flat bands. For the lowest LL and ideal topological flat bands in moir\'e systems, a common ingredient is that these flat bands can all be reduced to the problem of finding the null space of a non-Hermitian operator $\mathcal{D}(\overline{\partial_z}, z ,\overline{z})$, with $\mathcal{D} \psi = 0$, where $\mathcal{D}$ does not contain any $\partial_z$. For LLs, this operator is simply $2i\overline{\partial_z} - A$, where $A$ is the gauge field, while for chiral TBG, $\mathcal{D}$ is a $2 \times 2$ matrix.

If the Bloch wavefunctions of the ideal flat band $\psi_\mathbf{k}$ can be factorized into the form $\psi_{\mathbf{k}} = f_{\mathbf{k}-\mathbf{k}_0}(z,\overline z) \psi_{\mathbf{k}_0}$, where $\psi_{\mathbf{k}_0}$ is the Bloch wavefunction at an arbitrary reference momentum point $\mathbf{k}_0$, the null-space condition implies that $\mathcal{D} \psi_{\mathbf{k}}=\mathcal{D} \psi_{\mathbf{k}_0}=0$. Consequently, the function $f_{\mathbf{k}-\mathbf{k}_0}$ must satisfy the equation
\begin{eqnarray}
	\mathcal{D}(f_{\mathbf{k}-\mathbf{k}_0}\psi_{\mathbf{k}_0}) - f_{\mathbf{k}-\mathbf{k}_0} \mathcal{D}\psi_{\mathbf{k}_0} = 0
 \label{eq:general:PDE}
\end{eqnarray}
It is easy to verify that this equation only involves derivatives of $f$, as terms proportional to $f$ (without derivatives) cancels out. Hence, this equation is effectively an equation for $g_\mathbf{k} =\overline{\partial_z}f_\mathbf{k}$. 
More precisely, if $\mathcal{D}$ is a $n\times n$ matrix,  Eq.~\eqref{eq:general:PDE} represents a set of $n$ homogeneous partial different equations of $g$
\begin{eqnarray}
	\tilde{\mathcal{D}} g_{\mathbf{k}}= 0,
 \label{eq:general:PDE:g}
\end{eqnarray}
where the operator $\tilde{\mathcal{D}}(\overline{\partial_z}, z ,\overline{z})$ does not contain any holomorphic derivative $\partial_z$.

For homogeneous partial differential equations, a trivial solution always exists: $g=0$. Since $g$ is the anti-holomorphic derivative of $f$, this trivial solution means that $f$ is holomorphic. Most of the well-known models of ideal flat bands, such as the lowest LL and the chiral limit of TBG, belong to this category.

However, beyond this trivial solution, homogeneous PDEs may also support nontrivial solutions. For the lowest LL or 
TBG, it can be proved that nontrivial solutions are not allowed. However, for generic moir\'e systems, such solutions are possible, thus providing an alternative pathway towards ideal flat bands. The model discussed above falls into this category. It is important to note that the PDE for $g$ [Eq.~\eqref{eq:general:PDE:g}] has the same structure as the PDE that a chiral flat band must satisfy. Therefore, we can define another moir\'e system with an effective Hamiltonian $\mathcal{H}_\text{eff}= \begin{pmatrix}
        0 & \tilde{\mathcal{D}}^\dagger\\
        \tilde{\mathcal{D}} & 0
    \end{pmatrix}$
If $\mathcal{H}_\text{eff}$ has chiral flat bands, their Bloch wavefunction provides a nontrivial solution for Eq.~\eqref{eq:general:PDE:g}. Should the flat band of $\mathcal{H}_{\text{eff}}$ resemble the lowest Landau level with a Chern number $|C|=1$, this solution for $g$ leads to an ideal flat band with $|C|=2$ when we revert to the original model, as exemplified above. If the flat bands of $\mathcal{H}_{\text{eff}}$ have a higher Chern number $|C|=n$, the original model then contains flat bands with $|C|=n+1$.


\textit{Band crossings of other orders, splitting of band crossings.}--Existing studies have found $C=1$ holomorphic ideal flat bands can arise from Dirac and quadratic band touchings, i.e., band touchings with dispersion $E \propto |\mathbf{k}|^n$ for $n=1$ or $2$. In this work, we demonstrate that $n=4$ band touchings produce a different type of ideal flat band with $C=2$. This naturally raises the question: can other types of ideal flat bands be obtained from band crossings with $n=3$ or $n>4$? At least for single-layer systems, we find that the answer is negative. 

What is unique about $n=1$, $2$ and $4$ lies in the fact that these band touchings are protected by symmetries (three-fold rotation $C_{3z}$, time-reversal, and chiral). In contrast, for $n=3$ or $n>4$, the band crossing generally splits into multiple lower-order band crossings with $n=1$ or $2$, when a moir\'e potential is introduced (see SM~\cite{SM2024}). If the degeneracy point splits, the eigenvalues of the Birman-Schwinger operator $T_n(\mathbf{k}; \mathbf{r})$ generally do not remain \(\mathbf{k}\)-independent~\footnote{The eigenvalues of \(T_n(\mathbf{k}; \mathbf{r})\) indicate the values of \(\alpha\) at which there is a zero mode at \(\mathbf{k}\). If the degeneracy at \(\mathbf{k} = \mathbf{0}\) splits into Dirac crossings, the eigenvalues of \(T_n(\mathbf{k}; \mathbf{r})\) indicate at which value of \(\alpha\) the Dirac crossing reaches wave vector \(\mathbf{k}\). Since the Dirac crossing reaches different \(\mathbf{k}\) at different values of \(\alpha\), the eigenvalues of \(T_n(\mathbf{k}; \mathbf{r})\) become \(\mathbf{k}\)-dependent. See also Supplemental Material~\cite{SM2024}.}. Hence, exact flat bands are generally not possible for any \(n \not\in \{2,4\}\)~\footnote{In the original paper~\cite{becker2022mathematics} using the Birman-Schwinger operator to find ``magic'' angles, zero modes appearing only at $\mathbf{k}=\mathbf{0}$ for generic angles was one of the requirements for exact flat bands to appear at some ``magic'' angle. See also SM~\cite{SM2024}.}.

\textit{Multiply degenerate exact flat bands.}--Before concluding our discussion, we present another model to demonstrate the diversity and rich physics achievable from this new family of exact flat bands. As shown in the appendix, by simply choosing a slightly different moir\'e potential, our model can give rise to four degenerate exact flat bands. Two of these bands have a total Chern number of $C=2$, while the other two have $C=-2$. The wavefunctions of these flat bands can also be analytically constructed, with detailed explanations provided in the appendix and SM~\cite{SM2024}.

\textit{Discussion.}--The existing examples of exact flat bands with ideal quantum geometry in moir\'e systems found in the literature rely on the holomorphic structure $\psi_{\mathbf{k}} = f_{\mathbf{k} - \mathbf{k}_0}(z) \psi_{\mathbf{k}_0}$, where $f_\mathbf{k}(z)$ is a holomorphic function in $z$ (up to a modification by adding multiple layers). This structure allows these wavefunctions to be ``adiabatically" connected to those of the lowest LL, interpreting these wavefunctions as describing electrons in a spatially varying magnetic field~\cite{ledwith2020fractional, crepel2024topologically}. The family of single-component wavefunctions described in this article cannot be written in this form, indicating that they are distinct from the lowest LL family, and thus do not correspond to wavefunctions induced by magnetic fields, whether homogeneous or spatially varying. However, these wavefunctions do satisfy ideal quantum geometry, unlike higher LLs~\cite{ozawa2021relations} or ``higher vortexable" states~\cite{fujimoto2024higher}. 

It will be interesting to study possible fractional states in this family of flat bands. Since the Bloch wavefunctions here are not holomorphic in $z$, the many-body wavefunctions of fractional states must also contain non-holomorphic structures, extending beyond the well-understood holomorphic-function-based many-body wavefunctions, such as Laughlin wavefunctions. Additionally, because many fundamental concepts of fractional states are built upon the structure of holomorphic functions and polynomial functions of $z$, these insights need to be revisited in these new flat band systems~\cite{dong2023many, wang2023origin}.

Finally, we mention an interesting coincidence. If we define the ratio between the first and lowest Landau levels as $\tilde{f} = \psi_1 / \psi_0$, it is easy to verify that this $\tilde{f}$ function exhibits the same properties as the non-holomorphic $f$ function we find: although $\tilde{f}$ is non-holomorphic, its anti-holomorphic first-order derivative $\tilde{g} = \overline{\partial_z} \tilde{f}$ obeys the same condition with our $g$ function: multiplying $\tilde{g}$ by an arbitrary holomorphic function yields another valid $\tilde{g}$.

\begin{acknowledgments}
\noindent \textit{Acknowledgements}.--This work was supported in part by Air Force Office of Scientific Research MURI FA9550-23-1-0334 and the Office of Naval Research MURI N00014-20-1-2479 and Award N00014-21-1-2770, and by the Gordon and Betty Moore Foundation Award N031710 (KS).
\end{acknowledgments}

\bibliographystyle{apsrev4-1}
\bibliography{ref}


\setcounter{equation}{0}  
\setcounter{figure}{0}
\makeatletter
\renewcommand \thesection{Appendix \@Alph\c@section}
\makeatother
\renewcommand{\theequation}{A\arabic{equation}}
\renewcommand{\thefigure}{A\arabic{figure}}
\begin{widetext}
\section{Multiply degenerate exact flat bands.}

\begin{figure}[t]
     \centering
\includegraphics[scale=1]{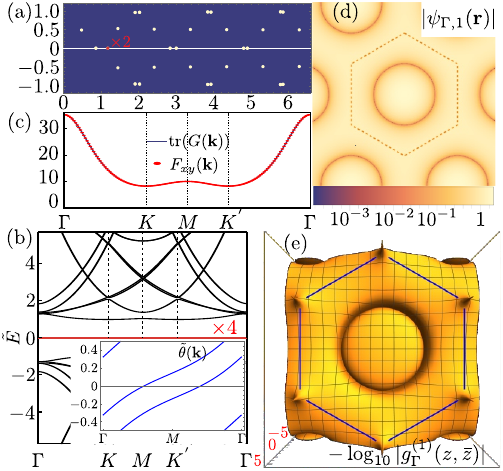}
     \caption{four degenerate exact flat bands from quartic band crossing point under Moir\'e potential with $p6mm$ symmetry. (a)  Magic values of $\alpha$. Each point represent a magic $\alpha$. For better visualization, here we showed the values of $\alpha^{1/4}$, where the horizontal and vertical axis are the real and imaginary part of $\alpha^{1/4}$.  
     (b) Band structure along the high symmetry path in the Moir\'e Brillouin zone using moir\'e potential shown in Eq.~\ref{eq:p6mm} at $\alpha=1.970$, which corresponds to the red point in (a). The vertical axis is normalized eigen-energies $\tilde{E}=\frac{E}{|\mathbf{G}_{i}|^2}$. There are four exact flat bands at energy $\tilde{E}=0$. Wilson loop spectrum $\tilde{\theta}({\mathbf{k}})=\frac{\theta({\mathbf{k}})}{2\pi}$ of the two sublattice polarized wavefunctions is shown at the right bottom corner of (b), which shows total Chern number $|C| = 2$.
     (c) Trace of non-Abelian Fubini-Study metric and non-Abelian Berry curvature of the two sublattice polarized wavefunctions along the high symmetry path in the Moir\'e Brillouin zone, which shows ideal non-Abelian trace condition.
     (d) Density plot of $|\psi_\Gamma(\mathbf{r})|$ which has no zero that can cancel the singularity of a hole. The boundary of the unit cell is plotted in red dash lines.
     (e) 3D plot of $-\log_{10} |g_\Gamma(z,\bar{z})|$, where peaks indicate zeros. Blue lines mark the boundary of the unit cell.
     }
     \label{fig:ExactFlatBand4}
\end{figure}

Here we choose a different moir\'e potential for the Hamiltonian defined in Eq.~\eqref{eq:H4}
\begin{equation}
	A(\mathbf{r}) = \frac{1}{2}\sum_{i=1}^3 e^{2\pi i (n-1)/3} \cos(\mathbf{G}_i\cdot\mathbf{r}).
    \label{eq:p6mm}
\end{equation}
The form of this potential means that Hamiltonian now has $\mathcal{C}_{6z}$ symmetry. Again, by evaluating the inverse of the eigenvalues of the Birman-Schwinger operator $T_4(\mathbf{k};\mathbf{r})$ at generic value of $\mathbf{k}$ (see {\color{black}Fig.~\ref{fig:ExactFlatBand4}(a)}), we find the ``magic'' $\alpha$ values for this potential. Interestingly, some of the eigenvalues of $T_4(\mathbf{k};\mathbf{r})$ are doubly degenerate now. This implies that there are two independent eigenmodes of $T_4(\mathbf{k};\mathbf{r})$, hence two independent zero modes of $\mathcal{D}_4(\mathbf{k};\mathbf{r})$. Therefore, there should be 4 exact flat bands in the spectrum of $\mathcal{H}_4(\mathbf{r})$, which we verified in {\color{black}Fig.~\ref{fig:ExactFlatBand4}(b)}.
But, how do we construct the wavefunctions at generic $\mathbf{k}$ in this case? To do so, we first evaluate $\psi_\Gamma(\mathbf{r})$. Note that now there two are exact flat band wavefunctions at $\Gamma$ on the same sublattice. One of them satisfies $\psi_{\Gamma,1}(\mathcal{C}_{3z}\mathbf{r}) =\psi_{\Gamma,1}(\mathbf{r})$, the other one satisfies $\psi_{\Gamma,2}(\mathcal{C}_{3z}\mathbf{r}) = e^{2\pi i/3}\psi_{\Gamma,2}(\mathbf{r})$. We choose the first one because it is always a zero mode even away from magic $\alpha$, and at magic $\alpha$, it acquires some special properties that allows for the construction of exact flat band wavefunctions at generic $\mathbf{k}$. Also, we show in the SM~\cite{SM2024} that starting from $\psi_{\Gamma,1}(\mathbf{r})$ constructing the of wavefunctions at generic $\mathbf{k}$, we can analytically continue them to $\mathbf{k} = 0$ to obtain $\psi_{\Gamma,1}(\mathbf{r})$. We plot $\psi_{\Gamma,1}(\mathbf{r})$ at the magic $\alpha$ value in {\color{black}Fig.~\ref{fig:ExactFlatBand4}(d)}. Although $\psi_{\Gamma,1}(\mathbf{r})$ has a ring of zeros around the center of the unit cell, the function $\psi_{\Gamma,1}(\mathbf{r})$ in the neighborhood of these zeros is not of type $F(z,\bar{z})z$ (where $F(z,\bar{z})$ is non-singular function), which we numerically checked. Hence, again, any lowest LL type exact flat band wavefunction construction of $f_\mathbf{k}(z)\psi_{\Gamma,1}(\mathbf{r})$, where $f_\mathbf{k}(z)$ is a holomorphic function of $z$, is not possible since such a $f_\mathbf{k}(z)$ necessarily has a pole of type $1/z$, which is not canceled in $f_\mathbf{k}(z)\psi_{\Gamma,1}(\mathbf{r})$ unless $\psi_{\Gamma,1}$ in the vicinity of its zeros is of type $F(z,\bar{z})z$. However, when we numerically evaluate $g_\Gamma^{(1)}(z,\bar{z})$ by solving Eq.~\eqref{Eq:geq}, we find that $g_\Gamma^{(1)}(z,\bar{z})$ has two zeros $z_0^{(1)}$ and $z_0^{(2)}$ at the two corners of the unit cell (shown in {\color{black}Fig.~\ref{fig:ExactFlatBand4}(e)}). Then, two independent functions $g_\mathbf{k}^{(1)}(z,\bar{z}) = h_\mathbf{k}(z;z_0^{(1)})g_\Gamma^{(1)}(z,\bar{z})$ and $g_\mathbf{k}^{(2)}(z,\bar{z}) = h_\mathbf{k}(z;z_0^{(2)})g_\Gamma^{(1)}(z,\bar{z})$ can be created, both of which satisfy Eq.~\eqref{Eq:geq}. From there, following the steps outlined in the previous example one can find the wavefunctions $\psi_\mathbf{k}^{(1)}(\mathbf{r})$ and $\psi_\mathbf{k}^{(2)}(\mathbf{r})$. 

Na\''ively, one may think that, each of these wavefunctions carries $C=2$, and in total the two wavefunctions carry Chern number $C = 4$. However, as was shown in~\cite{sarkar2023symmetry}, even though $g_\mathbf{k}^{(1)}(z,\bar{z})$ and $g_\mathbf{k}^{(2)}(z,\bar{z})$ are independent, they are not orthogonal. Upon orthogonalization, the two new functions can be taken as $\tilde{g}_\mathbf{k}^{(1)}(z,\bar{z}) = g_\mathbf{k}^{(1)}(z,\bar{z})$ and $\tilde{g}_\mathbf{k}^{(2)}(z,\bar{z}) = g_\mathbf{k}^{(2)}(z,\bar{z}) - \frac{\langle g_\mathbf{k}^{(1)}|g_\mathbf{k}^{(2)}\rangle}{\langle g_\mathbf{k}^{(1)}|g_\mathbf{k}^{(1)}\rangle}g_\mathbf{k}^{(1)}(z,\bar{z})$. As was shown in~\cite{sarkar2023symmetry}, $\tilde{g}_\mathbf{k}^{(2)}(z,\bar{z})$ is topologically trivial. Now, when we evaluate $f_\mathbf{k}^{(i)}$ by integrating $\tilde{g}_\mathbf{k}^{(i)}$, we find a singularity of $1/k$ type for $f_\mathbf{k}^{(1)}$ because $\tilde{g}_{\mathbf{k}=\mathbf{0}}^{(1)}(\mathbf{G}=\mathbf{0})\neq 0$, which makes the Chern number of $f_\mathbf{k}^{(1)}$ to be $C = 2$ (same as what was discussed between Eqs.~\eqref{eq:fk1} and~\eqref{eq:fk}). However, $\tilde{g}_{\mathbf{k}=\mathbf{0}}^{(2)}(\mathbf{G}=\mathbf{0}) = 0$ as we show in SM~\cite{SM2024}, and hence, there is no singularity in $f_\mathbf{k}^{(2)}$. So, its Chern number is $C=0$ (since $\tilde{g}_\mathbf{k}^{(2)}$ has $C=0$). Therefore, the total Chern number of the two bands together is $C=2$. This is verified in total winding of 2 of the Wilson loop spectrum in the inset of Fig.~\ref{fig:ExactFlatBand4}(b). Notice that this total Chern number intuitively matches with our expectation from the winding of the band crossing that we started with. We know that a winding of $n$ around a band crossing gives Berry phase of $n\pi$ around it, hence for quartic band crossing gives Berry phase of $4\pi$, which corresponds to Chern number $C=2$, and this argument is independent of the degeneracy of the flat bands. Furthermore, since $\mathcal{D}_4(\mathbf{k};\mathbf{r})$ is holomorphic in $k=k_x+ik_y$, and zero modes of $\mathcal{D}_4(\mathbf{k};\mathbf{r})$ are isolated from other eigenmodes of it (otherwise there would be other bands crossing energy $E = 0$), the zero modes of $\mathcal{D}_4(\mathbf{k};\mathbf{r})$ depend holomorphically on $k=k_x+ik_y$ (this is due to~\cite{kato2013perturbation} Chap. VII, Theorem 1.7). Hence, zero modes of $\mathcal{D}_4(\mathbf{k};\mathbf{r})$ (which are exact flat band wavefunctions) have ideal quantum geometry~\cite{ledwith2020fractional, claassen2015position}, as is verified numerically in {\color{black}Fig.~\ref{fig:ExactFlatBand4}(c)}.
\end{widetext}
\clearpage
\onecolumngrid
\vspace{10cm}

\makeatletter
\renewcommand \thesection{S-\@arabic\c@section}
\renewcommand\thetable{S\@arabic\c@table}
\renewcommand \thefigure{S\@arabic\c@figure}
\renewcommand \theequation{S\@arabic\c@equation}
\makeatother
\setcounter{equation}{0}  
\setcounter{figure}{0}  
\setcounter{section}{0}  
\counterwithin{figure}{section} 
{
    \center \bf \large 
    Supplemental Material\vspace*{0.1cm}\\ 
    \vspace*{0.0cm}
}
\maketitle

\section{Wave functions and topology of multiple flat bands on each sublattice}
In this section, we construct the wave functions for the case considered in Fig.~2 of the main text, and show that the total Chern number of the exact flat band wavefunctions polarized on the same sublattice is $C=2$.

Note that now there are two exact flat band wavefunctions at $\Gamma$ on the same sublattice. One of them satisfies $\psi_{\Gamma,1}(\mathcal{C}_{3z}\mathbf{r}) =\psi_{\Gamma,1}(\mathbf{r})$, the other one satisfies $\psi_{\Gamma,2}(\mathcal{C}_{3z}\mathbf{r}) = e^{2\pi i/3}\psi_{\Gamma,2}(\mathbf{r})$. We choose the first one because it is always a zero mode even away from magic $\alpha$, and at magic $\alpha$, it acquires some special properties that allows for the construction of exact flat band wavefunctions at generic $\mathbf{k}$. Also, we show below that starting from $\psi_{\Gamma,1}(\mathbf{r})$ constructing the of wavefunctions at genertic $\mathbf{k}$, we can analytically continue them to $\mathbf{k} = 0$ to obtain $\psi_{\Gamma,1}(\mathbf{r})$. 

\begin{figure}[h]
     \centering
\includegraphics[scale=2]{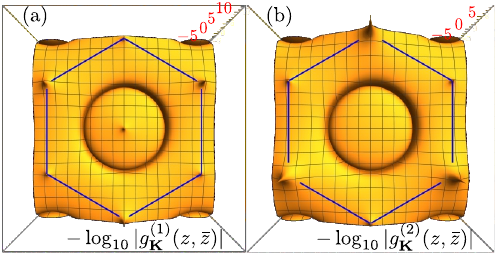}
     \caption{
     Numerical verification of construction of $g_\mathbf{k}^{(i)}(z,\bar{z})$ in Eq.~\eqref{eq:g12}. (a-b) 3D plot of $-\log_{10} |g^{(1)}_{\mathbf{K}}(z,\bar{z})|$ and $-\log_{10} |g^{(2)}_{\mathbf{K}}(z,\bar{z})|$, respectively. Here peaks indicate zeros. The boundary of the unit cell is plotted in Blue lines. We know from Fig.~2(e) of the main text that $g_\Gamma^{(1)}(z,\bar{z})$ has two zeros at positions $\mathbf{r}_0^{(1)} = (\mathbf{a}_1+2\mathbf{a}_2)/3$ and $\mathbf{r}_0^{(2)} = -(\mathbf{a}_1+2\mathbf{a}_2)/3$, where $\mathbf{a}_1 = a(1,0)$ and $\mathbf{a}_2 =a(-1,\sqrt{3})/2$. In the neighborhood of both of these zeros, we find numerically that $g_\Gamma^{(1)}(\mathbf{r})$ has the form $g_\Gamma^{(1)}(\mathbf{r}_0^{(i)}+(x,y))\sim z\bar{z}$, where $z=x+iy$. Since $h_\mathbf{k}(z+z_0^{(i)};z_0^{(i)})\sim 1/z$, $g_\mathbf{k}^{(i)}(\mathbf{r})$ should still have a zero at $\mathbf{r}_0^{(i)}$, but in the neighborhood of the zero, $g_{\mathbf{k}}^{(i)}(\mathbf{r})$ should have the form $g_{\mathbf{k}}^{(i)}(\mathbf{r}_0^{(i)}+(x,y))\sim \bar{z}$. However, since $h_\mathbf{k}(z;z_0^{(i)})$ also has a zero at $\mathbf{r}_0^{(i)}-\frac{\sqrt{3}a^2}{4\pi}\hat{z}\times \mathbf{k}$, $g_{\mathbf{k}}^{(i)}(\mathbf{r})$ should have a zero at this position. From the above arguments, we learn the following: when $\mathbf{k} = \frac{4\pi}{3a}(1,0)$ ($K$ point, the corner of the BZ), $g_K^{(1)}(z,\bar{z})$ should have a zero of type $\bar{z}$ at $(\mathbf{a}_1+2\mathbf{a}_2)/3$, a zero of type $z\bar{z}$ at $-(\mathbf{a}_1+2\mathbf{a}_2)/3$, and a zero of type of type $z$ at the center (with coordinate $\mathbf{0}$) of the unit cell. These positions of zeros can be seen in (a). Following the same logic, on the other hand, $g_K^{(2)}(z,\bar{z})$ should have a zero of type $\bar{z}$ at $-(\mathbf{a}_1+2\mathbf{a}_2)/3$, a zero of type $z^2\bar{z}$ at $(\mathbf{a}_1+2\mathbf{a}_2)/3$. These positions of zeros can be seen in (b).
     }
     \label{fig:s1}
\end{figure}

Starting from $\psi_{\Gamma,1}(\mathbf{r})$ in Fig.~2(d), when we numerically evaluate $g_\Gamma(z,\bar{z})$ by solving {\color{black}Eq.~(4)}, we find that $g_\Gamma^{(1)}(z,\bar{z})$ has two zeros $z_0^{(1)}$ and $z_0^{(2)}$ at the two corners of the unit cell (shown in {\color{black}Fig. 2(e) in the main text}). Then, two independent functions 
\begin{equation}
\label{eq:g12}
    \begin{split}
        g_\mathbf{k}^{(1)}(z,\bar{z}) = h_\mathbf{k}(z;z_0^{(1)})g_\Gamma^{(1)}(z,\bar{z}),\\
        g_\mathbf{k}^{(2)}(z,\bar{z}) = h_\mathbf{k}(z;z_0^{(2)})g_\Gamma^{(1)}(z,\bar{z})
    \end{split}
\end{equation}
can be created, where
\begin{equation}
\label{eq:hk}
    h_\mathbf{k}(z;z_0^{(i)}) = e^{\frac{i}{2} (\bar{k}z+k\bar{a}_1 z/a_1)}\frac{\vartheta\left(\frac{z-z_0^{(i)}}{a_1}-\frac{k}{b_2},\tau\right)}{\vartheta\left(\frac{z-z_0^{(i)}}{a_1},\tau\right)},
\end{equation}
where $a_1 = (\mathbf{a}_{1})_x+i(\mathbf{a}_{1})_y$ is the complexified moir\'e lattice vector $\mathbf{a}_1 = a(1,0)$, $b_2 = (\mathbf{b}_{2})_x+i(\mathbf{b}_{2})_y$ is the complexified moir\'e reciprocal lattice vector $\mathbf{b}_2 = 4\pi/\sqrt{3}a(0,1)$, $k = k_x+ik_y$, $\tau = e^{2\pi i/3}$, $\vartheta(z,\tau) = -i\sum_{n=-\infty}^\infty(-1)^n e^{\pi i \tau(n+1/2)^2+\pi i(2n+1)z}$ is the Jacobi theta function~\cite{TataI,ledwith2020fractional}. Both of the functions in Eq.~\eqref{eq:g12} satisfy {\color{black}Eq.~(4)} of the main text. 
{\color{black}To verify that this construction is indeed correct, we show the numerically obtained $g_{K}^{(i)}(z,\bar{z})$ at the corner of the Brillouin zone ($K$ point) in Fig.~\ref{fig:s1}. Indeed Eq.~\ref{eq:g12} can predict the positions of the zeros in these two functions correctly, as discussed in the caption of Fig.~\ref{fig:s1}.}

From $g_\mathbf{k}^{(i)}(z,\bar{z})$, following the steps outlined in the previous example one can find the wavefunctions $\psi_\mathbf{k}^{(1)}(\mathbf{r})$ and $\psi_\mathbf{k}^{(2)}(\mathbf{r})$. However, there are some subtleties here which dictate the topology of the wavefunctions.

\begin{figure}[h]
     \centering
\includegraphics[scale=2]{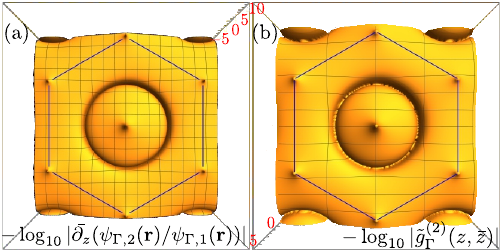}
     \caption{
     (a) 3D plot of $-\log_{10} |\bar{\partial_z}(\psi_{\Gamma,2}(\mathbf{r})/\psi_{\Gamma,1}(\mathbf{r}))|$ obtained numerically. 
     (b) 3D plot of $\tilde{g}_{\Gamma}^{(2)}(z,\bar{z})$ obtained from the construction in Eq.~\eqref{eq:g2gamma}.
     }
     \label{fig:s2}
\end{figure}

Na\''ively, one may think that, each of these wavefunctions carries $C=2$, and in total the two wavefunctions carry Chern number $C = 4$. However, as was shown in~\cite{sarkar2023symmetry}, even though $g_\mathbf{k}^{(1)}(z,\bar{z})$ and $g_\mathbf{k}^{(2)}(z,\bar{z})$ are independent, they are not orthogonal. Upon orthogonation, the two new function can be taken as 
\begin{equation}
\label{eq:g2k}
    \begin{split}
        \tilde{g}_\mathbf{k}^{(1)}(z,\bar{z}) &= g_\mathbf{k}^{(1)}(z,\bar{z}),\\
        \tilde{g}_\mathbf{k}^{(2)}(z,\bar{z}) &= g_\mathbf{k}^{(2)}(z,\bar{z}) - \frac{\langle g_\mathbf{k}^{(1)}|g_\mathbf{k}^{(2)}\rangle}{\langle g_\mathbf{k}^{(1)}|g_\mathbf{k}^{(1)}\rangle}g_\mathbf{k}^{(1)}(z,\bar{z}).
    \end{split}
\end{equation}
 As was shown in~\cite{sarkar2023symmetry}, $\tilde{g}_\mathbf{k}^{(2)}(z,\bar{z})$ is topologically trivial (we discuss why that is the case below). Now, when we evaluate $f_\mathbf{k}^{(i)}$ by integrating $\tilde{g}_\mathbf{k}^{(i)}$, we find a singularity of $1/k$ type for $f_\mathbf{k}^{(1)}$ because $\tilde{g}_{\mathbf{k}=\mathbf{0}}^{(1)}(\mathbf{G}=\mathbf{0})\neq 0$, which makes the Chern number of $f_\mathbf{k}^{(1)}$ to be $C = 2$ (same as what was discussed between {\color{black}Eqs.~(6) and~(7) of main text}). However, $\tilde{g}_{\mathbf{k}=\mathbf{0}}^{(2)}(\mathbf{G}=\mathbf{0}) = 0$ as we show below, and hence, there is no singularity in $f_\mathbf{k}^{(2)}$. So, its Chern number is $C=0$ (since $\tilde{g}_\mathbf{k}^{(2)}$ has $C=0$).
 First notice that the expression of $\tilde{g}_\mathbf{k}^{(2)}(z,\bar{z})$ in  Eq.~\eqref{eq:g2k} implies that $\tilde{g}_{\mathbf{k}=\mathbf{0}}^{(2)}(z,\bar{z}) = 0$ since  $h_{\mathbf{k}=\mathbf{0}}(z;z_0^{(1)}) = h_{\mathbf{k}=\mathbf{0}}(z;z_0^{(2)}) = 1$. Therefore, we have to find $\tilde{g}_{\mathbf{k}=\mathbf{0}}^{(2)}(z,\bar{z})$ by doing an analytic continuation from nonzero $\mathbf{k}$. To this end, let us first define the unit cell periodic part of $h_\mathbf{k}(z;z_0^{(i)})$ as
 \begin{equation}
     \tilde{h}_\mathbf{k}(z;z_0^{(i)}) = e^{-i\mathbf{k}\cdot\mathbf{r}}h_\mathbf{k}(z;z_0^{(i)}) = e^{\frac{i}{2} k(\bar{a}_1 z/a_1-\bar{z})}\frac{\vartheta\left(\frac{z-z_0^{(i)}}{a_1}-\frac{k}{b_2},\tau\right)}{\vartheta\left(\frac{z-z_0^{(i)}}{a_1},\tau\right)} = e^{-i k(\mathbf{b}_2\cdot\mathbf{r})/b_2}\frac{\vartheta\left(\frac{z-z_0^{(i)}}{a_1}-\frac{k}{b_2},\tau\right)}{\vartheta\left(\frac{z-z_0^{(i)}}{a_1},\tau\right)},
 \end{equation}
 which is a holomorphic function of $k$. Around $\mathbf{k} =\mathbf{0}$, $h_\mathbf{k}(z;z_0^{(i)})$ can be expanded as
 \begin{equation}
     h_\mathbf{k}(z;z_0^{(i)}) = 1 + \tilde{h}_\mathbf{0}'(z;z_0) k +i(k \bar{z}+\bar{k} z)/2 +\mathcal{O}(k^2),
 \end{equation}
 where $\tilde{h}_\mathbf{0}'(z;z_0) \equiv [\partial_k \tilde{h}_\mathbf{0}(z;z_0)]|_{\mathbf{k} = \mathbf{0}}=\frac{1}{2}[(\partial_{k_x}-i\partial_{k_y})\tilde{h}_\mathbf{k}(z;z_0)]|_{\mathbf{k} = \mathbf{0}}$ and we used $h_{\mathbf{k}=\mathbf{0}}(z;z_0^{(i)}) = \tilde{h}_{\mathbf{k}=\mathbf{0}}(z;z_0^{(i)}) = 1$. Plugging this into the expression of $\tilde{g}_\mathbf{k}^{(2)}(z,\bar{z})$, we obtain
 \begin{equation}
 \label{eq:g2gammak}
     \tilde{g}_\mathbf{k}^{(2)}(z,\bar{z}) =k(\tilde{h}_\mathbf{0}'(z;z_0^{(2)})-\tilde{h}_\mathbf{0}'(z;z_0^{(1)})-\langle g_{\Gamma}^{(1)}|(\tilde{h}_\mathbf{0}'(z;z_0^{(2)})-\tilde{h}_\mathbf{0}'(z;z_0^{(1)}))g_{\Gamma}^{(1)}\rangle)g_{\Gamma}^{(1)}(z,\bar{z})+\mathcal{O}(k^2).
 \end{equation}
 Continuing this function to $\mathbf{k}=\mathbf{0}$, we get
 \begin{equation}
 \label{eq:g2gamma}
     \tilde{g}_{\mathbf{k}=\mathbf{0}}^{(2)}(z,\bar{z}) =(\tilde{h}_\mathbf{0}'(z;z_0^{(2)})-\tilde{h}_\mathbf{0}'(z;z_0^{(1)})-\langle g_{\Gamma}^{(1)}|(\tilde{h}_\mathbf{0}'(z;z_0^{(2)})-\tilde{h}_\mathbf{0}'(z;z_0^{(1)})|g_{\Gamma}^{(1)}\rangle)g_{\Gamma}^{(1)}(z,\bar{z}).
 \end{equation}
 Note that this procedure was used in~\cite{sarkar2023symmetry} successfully to obtain wavefunctions for multiply degenerate flat bands. We plot $\tilde{g}_{\mathbf{k}=\mathbf{0}}^{(2)}(z,\bar{z}) \equiv \tilde{g}_{\Gamma}^{(2)}(z,\bar{z})$ in {\color{black}Fig.~\ref{fig:s2}(b)}. Several comments are in order.
 \begin{enumerate}
     \item Note that $\tilde{g}_{\Gamma}^{(2)}(z,\bar{z})$ obtained this way should be exactly the same as what we would get if we evaluate $\bar{\partial_z}(\psi_{\Gamma,2}(\mathbf{r})/\psi_{\Gamma,1}(\mathbf{r}))$. We plot the numerically obtained $\bar{\partial_z}(\psi_{\Gamma,2}(\mathbf{r})/\psi_{\Gamma,1}(\mathbf{r}))$ in {\color{black}Fig.~\ref{fig:s2}(a)}. The fact that the two plots in {\color{black}Fig.~\ref{fig:s2}} match well proves that the construction in Eq.~\eqref{eq:g2gamma} is correct. 
     \item Notice that the singularity $\tilde{g}_\mathbf{k}^{(2)}\propto k$ in Eq.~\eqref{eq:g2gammak} has opposite winding than the singularity in $f_\mathbf{k}(z) \propto 1/k$ shown earlier in the main text between {\color{black}Eqs.~(6) and (7)}. Hence, unlike the singularity $f_\mathbf{k}(z)$ which increases the Chern number by 1, the singularity in $\tilde{g}_\mathbf{k}^{(2)}$ decreases the Chern number by 1, this is why the Chern number carried by $\tilde{g}_\mathbf{k}^{(2)}$ is zero.
     \item We numerically find that the function $\tilde{g}_{\mathbf{k}=\mathbf{0}}^{(2)}(z,\bar{z})$ under $\mathcal{C}_{3z}$ transforms as $\tilde{g}_{\mathbf{k}=\mathbf{0}}^{(2)}(\mathcal{C}_{3z}z,\overline{\mathcal{C}_{3z}z}) = {\color{black}e^{-2\pi i/3}}\tilde{g}_{\mathbf{k}=\mathbf{0}}^{(2)}(z,\bar{z})$. This transformation property is also intuitively clear since we know that $g_{\Gamma}^{(1)}$ transforms as a scalar under $\mathcal{C}_{3z}$ ($\tilde{g}_{\Gamma}^{(1)}(\mathcal{C}_{3z}z,\overline{\mathcal{C}_{3z}z}) = \tilde{g}_{\Gamma}^{(1)}(z,\bar{z})$), $h_{\mathbf{k}=\mathbf{0}}(z;z_0^{(i)})=1$, and under $\mathcal{C}_{3z}$ the partial derivative with respect to complex $k$ transforms to $\partial_k \rightarrow {\color{black}e^{-2\pi i/3}}\partial_k$. Since $\tilde{g}_{\mathbf{k}=\mathbf{0}}^{(2)}(\mathcal{C}_{3z}z,\overline{\mathcal{C}_{3z}z}) = {\color{black}e^{-2\pi i/3}}\tilde{g}_{\mathbf{k}=\mathbf{0}}^{(2)}(z,\bar{z})$, it necessarily has zero average value in the unit cell, or in other words $\tilde{g}_{\mathbf{k}=\mathbf{0}}^{(2)}(\mathbf{G}=\mathbf{0}) = 0$.
     \item Since $\tilde{g}_{\mathbf{k}=\mathbf{0}}^{(2)}(\mathbf{G}=\mathbf{0}) = 0$, we can directly use {\color{black}Eq.~(6)} to obtain $f_{\mathbf{k}=\mathbf{0}}^{(2)}$. Multiplying $f_{\mathbf{k}=\mathbf{0}}^{(2)}$ by $\psi_{\Gamma,1}(\mathbf{r})$, we can obtain $\psi_{\Gamma,2}(\mathbf{r})$ as promised.
 \end{enumerate}
 \section{Why does the quartic band crossing not split under the addition of moir\'e potential?}
We know from representation theory that in 2D spinless systems only linear (at a $\mathcal{C}_{2z}\mathcal{T}$ or $\mathcal{I}\mathcal{T}$ ($\mathcal{I}$ is inversion) symmetric $\mathbf{k}$ point) and quadratic (at a time reversal invariant $\mathbf{k}$ point which is also $\mathcal{C}_{n'z}$ ($n'\geq3$) invariant) band crossings are stable. Then one may wonder why the quartic band crossing in our system does not split into Dirac crossings under the application of small moir\'e potential. The reason behind this comes from two things: (i) the constraints from chiral symmetry $\mathcal{S}$, time reversal symmetry $\mathcal{T}$ and three fold rotation symmetry $\mathcal{C}_{3z}$  on the system, (ii) the holomorphic dependence of $\mathcal{D}_4(\mathbf{k};\mathbf{r})$ on the parameter $\alpha$. We discuss this below. 

Even though we want to show that the band crossing at energy $E=0$ at $\mathbf{k} = \mathbf{0}$ does not split for quartic band crossing with the addition of the moir\'e potential, our procedure will also show why the band crossing of order $n=3$ and $n\geq 5$ splits into crossings of lower orders. Hence, we keep the discussion general for the first few steps.

\begin{figure}[h]
     \centering
\includegraphics[scale=2]{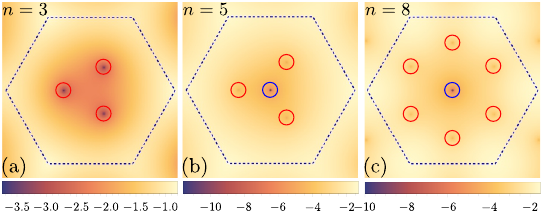}
     \caption{
     Density plot of the lowest band energy $\log_{10}|E_1(\mathbf{k})|$ in the moir\'e Brillouin zone (blue dashed lines show the edge of the Brillouin zone) showing splitting of band crossings 
     of different orders for small values of amplitude $\alpha$ of the moir\'e potential $A(\mathbf{r})$.
     Here, (a-c) correspond to higher order n crossing with $n=3,5,8$ respectively. The forms of the moir\'e potentials in (a), (b) and (c) are $A(\mathbf{r}) = \sum_{i=1}^3 e^{i\mathbf{-G}_i\cdot\mathbf{r}}$, $A(\mathbf{r}) = \sum_{i=1}^3 e^{2\pi i (n-1)/3} e^{-i\mathbf{G}_i\cdot\mathbf{r}}$, $A(\mathbf{r}) = \sum_{i=1}^3 e^{2\pi i (n-1)/3} e^{-i\mathbf{G}_i\cdot\mathbf{r}}$, respectively, such that system has $\mathcal{C}_{3z}$ and $\mathcal{M}_y$. The dark points inside the red (blue) circles correspond to linear Dirac (quadratic) crossings.
     }
     \label{fig:s3}
\end{figure}

We know that $\mathcal{D}_n(\mathcal{C}_{3z}\mathbf{k};\mathcal{C}_{3z}\mathbf{r}) = e^{2\pi i n/3} \mathcal{D}(\mathbf{k};\mathbf{r})$. When $\alpha = 0$,  $\mathcal{D}_n(\mathbf{k};\mathbf{r}) = (k_x+ik_y)^n$. At $\alpha = 0$, if we insist on moir\'e periodicity, there are degeneracies at the edge of the Brillouin zone at finite nonzero energies due to zone folding. For perturbatively small $\alpha$'s, band gaps will open at finite nonzero energy at the edge of the moir\'e Brillouin zone. Since we are interested in the splitting (or lack thereof) at zero energy at the zone center, we can project the Hamiltonian to the two bands closest $E = 0$. Since the moir\'e potential does not break $\mathcal{C}_3z$ or $\mathcal{S}$, the $\mathbf{k}\cdot\mathbf{p}$ Hamiltonian $\mathcal{H}_\text{eff}$ near $\mathbf{k} = \mathbf{0}$ should have the form
\begin{equation}
    \mathcal{H}_\text{eff} = \begin{pmatrix}
        0 & \mathcal{D}_{\text{eff},n}^\dagger(\mathbf{k})\\
        \mathcal{D}_{\text{eff},n}(\mathbf{k}) & 0
    \end{pmatrix},
\end{equation}
where
\begin{equation}
    \mathcal{D}_\text{eff}(\mathbf{k}) = (k_x+i k_y)^n + f(k_x+i k_y),
\end{equation}
where $f$ is some holomorphic function of $k_x+ik_y$. The function $f$ needs to be holomorphic in $k_x+ik_y$ because the original $\mathcal{D}(\mathbf{k};\mathbf{r})$ is holomorphic; hence a faithful $\mathcal{D}_{\text{eff},n}$ should also be holomorphic. Now, since $\mathcal{D}_{\text{eff},n}$ has to also satisfy $\mathcal{D}_{\text{eff},n}(\mathcal{C}_{3z}\mathbf{k}) = e^{2\pi i n/3}\mathcal{D}_{\text{eff},n}(\mathbf{k})$, $f$ must satisfy $f(e^{2\pi i/3}(k_x+i k_y)) = e^{2\pi i n/3}f(k_x+i k_y)$. For different values of $n$, $f(k_x+i k_y)$ to the lowest order in $k$ are
\begin{equation}
    f(k_x+i k_y) =\begin{cases}
        c+\mathcal{O}((k_x+i k_y)^3), & n = 3,\\
        c(k_x+ik_y) + \mathcal{O}((k_x+i k_y)^4), & n = 4,\\
        c(k_x+ik_y)^2 + \mathcal{O}((k_x+i k_y)^5), & n = 5,\\
        c(k_x+ik_y)^{n \text{ mod } 3} + \mathcal{O}((k_x+i k_y)^{3+(n \text{ mod } 3)}), & \text{for any $n$},\\
    \end{cases}
\end{equation}
where $c$ is some complex number that is a function of $\alpha$ such that $c(\alpha=0) = 0$. Note that $c$ is a real number if the system has mirror symmmetry. For order $n$ band crossing to split, $\mathcal{D}_{\text{eff},n}$ must be zero at some nonzero $\mathbf{k}$. For $n=3$ or $n=5$, setting $\mathcal{D}_\text{eff}(\mathbf{k})=0$, we find that there are three $\mathbf{k} \neq 0$ solutions: $k_x+i k_y = (-c)^{1/3},e^{2\pi i/3}(-c)^{1/3},e^{4\pi i/3}(-c)^{1/3}$. This is exactly what we see in Figs.~\ref{fig:s3}(a,b). However, for $n=4$, there is one more symmetry: time reversal. Time reversal symmetry prohibits the term $c(k_x+ik_y)$ in $\mathcal{D}_{\text{eff},4}$. Hence, quartic band crossing cannot split.

 \section{Splitting of band crossings of different orders}

\section{More on the Birman-Schwinger operator $T_n(\mathbf{k};\mathbf{r})$}
To find if there is a ``magic'' value of $\alpha$ at which exact flat bands appear in this model, we utilize a method introduced in~\cite{becker2022mathematics}. We look at the structure of Bloch Hamiltonian $\mathcal{H}_n(\mathbf{k};\mathbf{r}) = e^{-i\mathbf{k}\cdot\mathbf{r}} \mathcal{H}_n(\mathbf{r})e^{i\mathbf{k}\cdot\mathbf{r}}$, whose eigenfunctions are periodic, where $\mathcal{H}_n(\mathbf{r})$ is a $2\times 2$ moir\' Hamiltonian with $n$-th order band crossing. The $\mathcal{H}_n(\mathbf{k};\mathbf{r})$ has off-diagonal term $\mathcal{D}_n(\mathbf{k};\mathbf{r}) = (2i\overline{\partial_z}-k)^n+\alpha A(\mathbf{r})$, where $k = k_x+ik_y$ is the complexified wave-vector. If there is an exact flat band, $\mathcal{D}_n(\mathbf{k};\mathbf{r})$ has zero modes for all $\mathbf{k}$. Writing $\mathcal{D}_n(\mathbf{k};\mathbf{r}) = (2i\overline{\partial_z}-k)^4+\alpha A(\mathbf{r}) = (2i\overline{\partial_z}-k)^n(\mathds{1}- \alpha T_4(\mathbf{k};\mathbf{r}))$ where we defined
\begin{equation}
    T_n(\mathbf{k};\mathbf{r}) = - (2i\overline{\partial_z}-k)^{-n}A(\mathbf{r}),
\end{equation}
this is the Birman-Schwinger operator~\cite{becker2022mathematics, becker2021spectral}. Since $(2i\overline{\partial_z}-k)^n$ is nonsingular when $\mathbf{k}$ is not a reciprocal lattice vector ($\mathbf{k}\neq m_1\mathbf{G}_1+m_2\mathbf{G}_2$, $m_i\in \mathds{Z}$) for periodic functions, any zero mode of $\mathcal{D}_n(\mathbf{k};\mathbf{r})$ at these non-special $\mathbf{k}$ values has to be a zero mode of  $(\mathds{1}- \alpha T_n(\mathbf{k};\mathbf{r}))$, and hence an eigen-mode of $T_n(\mathbf{k};\mathbf{r})$ with eigenvalue $1/\alpha$. Therefore, if the eigenvalues $\eta_\mathbf{k}$ of $T_n(\mathbf{k};\mathbf{r})$ are independent of $\mathbf{k}$, then the ``magic'' $\alpha$'s at which the exact flat bands appear are $\alpha = 1/\eta_\mathbf{k}$.
\subsection{When are $T_n(\mathbf{k};\mathbf{r})$ eigenvalues independent of $\mathbf{k}$?}
The eigenvalues of $T_n(\mathbf{k};\mathbf{r})$ just indicate the value of $\alpha$ at which there is a zero mode at $\mathbf{k}$. There are two cases that we need consider here:
\begin{enumerate}
    \item \underline{\textit{if the band crossing at $\mathbf{k} = \mathbf{0}$ does not split under the addition of moir\'e potential:}} This is the case for $n \in \{2,4\}$. In this case, we prove that the eigenvalues of $T_n(\mathbf{k};\mathbf{r})$ are independent of $\mathbf{k}$ by contradiction (the mathematically rigorous proof can be found in~\cite{becker2022mathematics}. here we give a physical argument). If $T_n(\mathbf{k};\mathbf{r})$ eigenvalues ($\eta_\mathbf{k}$) are dependent of $\mathbf{k}$, we can set $\alpha=\eta_\mathbf{k}$, then there would be an isolated zero mode at $\mathbf{k}$ for this value of $\alpha$. Around this zero mode, there would nonzero winding of the Berry phase. That would imply the total winding of the system around the Brillouin zone has changed form $n$ (the value we started with), which is impossible. This implies that $T_n(\mathbf{k};\mathbf{r})$ eigenvalues are independent of $\mathbf{k}$.

    \item \underline{\textit{if the band crossing at $\mathbf{k} = \mathbf{0}$  splits under the addition of moir\'e potential:}} This is the case for  $n= 3$ or $n\geq 5$ (as exemplified in Fig.~\ref{fig:s3}). In this case, the eigenvalues of $T_n(\mathbf{k};\mathbf{r})$ indicate at what value of $\alpha$ the Dirac crossing reaches wave vector $\mathbf{k}$. Since the Dirac crossing reaches different $\mathbf{k}$ at different value of $\alpha$, the eigenvalues of $T_n(\mathbf{k};\mathbf{r})$ are now $\mathbf{k}$ dependent.
\end{enumerate}
\end{document}